\documentclass[10pt]{forma}

\usepackage{graphicx}
\def\hho{H$_2$O~}
\def\Martin{Mart\'{\i}n}
\def\Bejar{B\'ejar}
\def\ee{\end{equation}}

\def\bea{\begin{eqnarray}}
\def\eea{\end{eqnarray}}

\def\etal{{\it et al.}}
\begin{document}
%%%%%%%%%%%%%%%%%%%%%%%%%%%%%%%%%%%%%%%%%%
\hspace{1.0cm} \parbox{15.0cm}{

\baselineskip = 15pt

\noindent {\bf ULTRACOOL DWARFS}
%%%%%%%%%%%%%%%%%%%%%%%%%%%%%%%%%%%%%%%%%%
\bigskip
\bigskip

\noindent {\bf Yakiv V. Pavlenko}
%%%%%%%%%%%%%%%%%%%%%%%%%%%%%%%%%%%%%%%%%%
\bigskip
\bigskip

\baselineskip = 9.5pt

$^1$\noindent {\small {\it Main Astronomical Observatory, NAS of Ukraine \\
27 Akademika Zabolotnoho Str., 03680 Kyiv, Ukraine}} \\
\noindent {\small {\it e-mail:}} {\tt yp@mao.kiev.ua}

%%%%%%%%%%%%%%%%%%%%%%%%%%%%%%%%%%%%%%%%%%
\baselineskip = 9.5pt \medskip

\medskip \hrule \medskip

\noindent

We present results of modeling of spectra of M-, L-, T-dwarfs.
Theoretical spectra are fitted to observed spectra to study the
main parameters of the low-mass objects beyond the bottom of Main
Sequence. Application of ``lithium'' and ``deuterium'' tests for
assessment of ultra-cool dwarfs are discussed.

\medskip \hrule \medskip

}
%%%%%%%%%%%%%%%%%%%%%%%%%%%%%%%%%%%%%%%%%%
\vspace{1.0cm}

\renewcommand{\thefootnote}{\ }

\footnotetext{\copyright~Yakiv V.Pavlenko 2004}

\renewcommand{\thefootnote}{\arabic{footnote}}

\baselineskip = 11.2pt
%%%%%%%%%%%%%%%%%%%%%%%%%%%%%%%%%%%%%%%%%%

\noindent {\small {\bf INTRODUCTION}}
\medskip

Population of ultracool (UC) dwarfs occupies the right-right-bottom quadrant
below the bottom of the Main sequence.
A lot of UC dwarfs was discovered after 1995
(see \cite{_Basri2000} and \cite{_BD2002} for reviews).
Basically,
we can define at least 3 different populations of ultracool dwarfs:

--- Low mass stars (LMS). Hydrogen burns in their core.

--- Brown dwarfs (BD). Hydrogen cannot burns in their core.
    Their existence were predicted by Kumar \cite{_Kumar1963a}, \cite{_Kumar1963b}.
    Later inverstigations
    show that lithium burns inside the brown dwarfs of $55M_j < M < 75M_j$
    (see \cite{_Chabrier2000a} for more details). Here $M_j$ is
    mass of Jupiter: 1$M_j = 0.001M_{\odot}$. First brown dwarfs Teide1 and Gl 229B
    were discovered by groups of Rebolo \cite{_Rebolo1995} and
    Nakajima \cite{_Nakajima1995}, respectively. Deuterium should be depleted in
    atmospheres of brown dwarfs.

--- Planets ($M < 13M_J$) preserve deuterium (and lithium) during their evolution
\cite{_Saumon1996}.

First spectral classifications of UC dwarfs were provided by
Kirkpatrick \etal \cite{_K99} and \Martin ~ \etal \cite{_M99}. Today
we can asses their spectra (see libraries of spectra on
\cite{_L2001} or \cite{_mountes2000}):

--- M-dwarfs (GJ406, VB10, VB8, etc). TiO dominates in their spectra.

--- L-dwarfs (GD169B, Kelu1,2MASS 0920+35, etc.). K and Na lines are the main features there
(\cite{_P1997},\cite{_P1998}), Ti and V atoms are bound into dust
particles.

--- T-dwarfs (Gl 229B, SDSS 0151, SDSS 1110, etc) -- infrared spectra show 
CH$_4$ lines.

--- planets (see list of discovered planets on web \cite{_planets},
and references therein). First confirmed discovery of planetary system 51 Peg
was carried out by Mayor \& Queloz \cite{_Mayor1995}
(see  Marcy et al. \cite{_Marcy1997}).

M-, -L,  -T dwarfs are of different effective temperatures and
masses. Still, ``the Main sequence'' for brown dwarfs and L-,
T-dwarfs forms the approximately horizontal line (Jupiter is on
the left side radii-masses plot, see \cite{_BL1993}) -- the
dependence of radii of UC dwarfs on mass is extremely weak due to
degeneracy of the gas in their cores. As result, sizes of old
brown dwarfs, L-dwarfs and Jupiter are comparatible.

As was noted by Zapatero Osorio ({\it private communication})
depending on age, T-dwarfs can be brown dwarfs (if they are old)  or
"planetary objects"  (masses below the deuterium burning limit, if
they are young). Hence, very young T-dwarfs do not burn deuterium.
Then, giant planets have been found by indirect techniques around stars.
Young objects a few times more massive than Jupiter have been
identified using direct imaging techniques. They are characterized by
ultracool atmospheres (L and T types). These objects are free-floating
in star-forming regions and very young clusters. This poses challenge
to current theories of stellar and planetary formation (see Proc. of
IAUS 211 \cite{_BD2002}).

Different UC dwarfs are of different structure as well:

--- inside the LMS we have core with hydrogen burning zone,

--- Brown dwarfs burn deuterium, the most massive BDs ($55M_J<M<75M_J$) burn
lithium within short time scales (see refs in \cite{_ZO2001}).

--- planets are only objects without any nuclear burning processes.
They preserve deuterium and lithium from times of their formation.

\newpage

\noindent {\small {\bf Models of formation of spectra of ultracool dwarfs}}
\medskip

\begin{figure}
\begin{center}
\includegraphics[width=7.0cm]{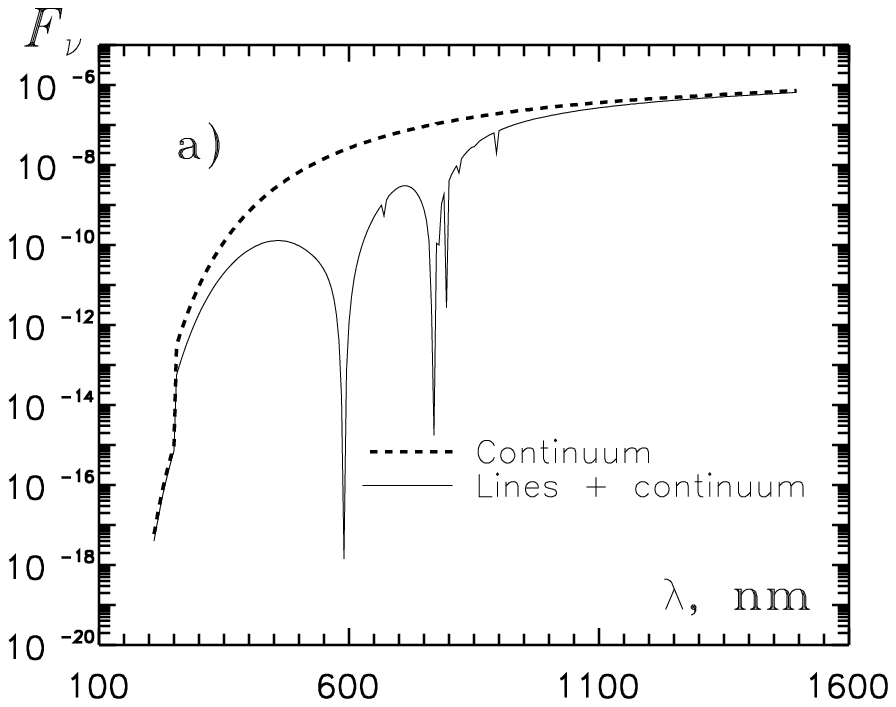}
\includegraphics[width=7.0cm]{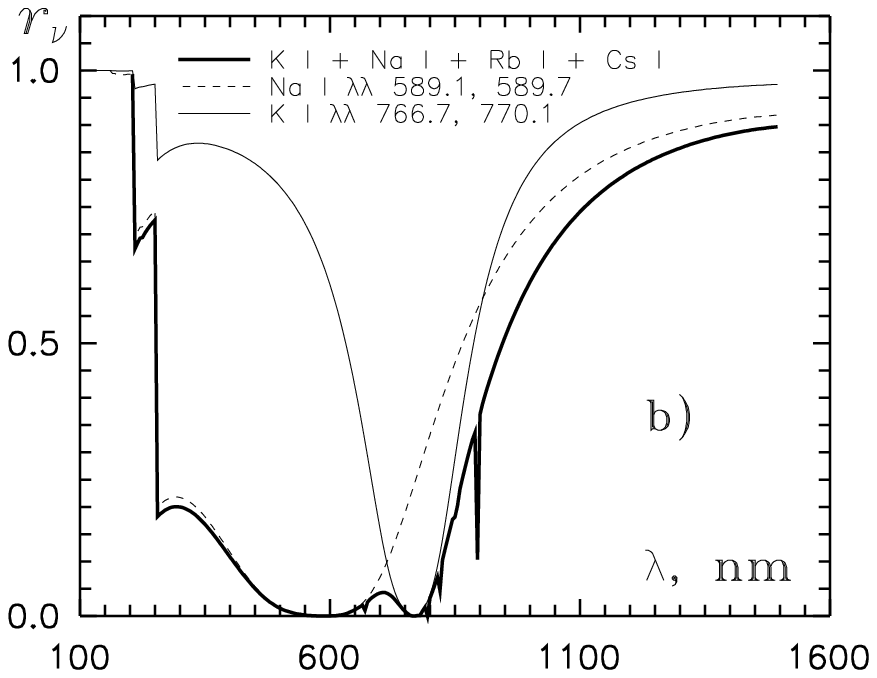}
\end{center}
\caption{Profiles of resonance lines of K I
($\lambda\lambda$ 766.6, 770.1 nm) and
Na I ($\lambda\lambda$ 589.1, 589.7 nm) computed
in the frame of collisional
broadening theory (van der Waals broadening)
for 1200/5.0 C-model atmosphere of Tsuji \cite{_Tsuji1998},
see \cite{_P2001} for more details.
\label{_KNa}}
\end{figure}

To model spectra and spectral energy distributions (SEDs) of ultracool dwarfs
we should account a few complicate processes which govern physical state
of their atmospheres:

\begin{itemize}

\item {\it Dust formation processes}.
Due to the low temperatures and high pressure regime some
molecular (and atomic) species are bound in different grain particles
(see \cite{_Tsuji1996}). Indeed, molecular bands of VO and TiO are weaker
in L-dwarf spectra in comparison with M-dwarfs.

\item {\it Damping of K and Na lines}. Resonance doublets of K and Na form
the most impressive features in spectra of L-dwarfs. Formally
computed equivalent widths of there lines can be of order a few
k\AA (see Fig. \ref{_KNa} and \cite{_P1998, _P2000} for more
details).

\item{\it Dust opacities}. Importance of account of dust opacities for
a procedure of numerical modelling of spectra of L- and T- dwarfs was
shown by Pavlenko ~\etal \cite{_P2000}. 
Basically, the problem of the dust opacities in L-dwarf
atmosphere is rather complicate -- we should account
absorption/scattering by particles of different composition, sizes,
orientations. Moreover, recent recearches provide some evidences of cloudy
structure of dust layers of L-dwarfs atmospheres (see materials of IAUS 211
\cite{_BD2002}).

\end{itemize}

%%%%%%%%%%%%%%%%%%%%%%%%%%%%%%%%%%%%%%%%%%
%%%%%%%%%%%%%%%%%%%%%%%%%%%%%%%%%%%%%%%%%%

\medskip
\noindent {\small {\bf Optical spectra: K and Na lines}}
\medskip

Resonance lines of Na I($\lambda\lambda$ 589.1, 589.7 nm) and K I
($\lambda\lambda$ 766.6, 770.1 nm) are very strong in spectra of UC
dwarfs \cite{_P1997a}, because majority of alkali atoms exists there as neutral
atoms. Na I resonance lines are stronger
--- in atmospheres of majority stars log N(Na) $>$ log N(K).

Lines of alkali metals observed in UC dwarfs spectra are pressure
broadened. Extremely strong broadening of K and Na resonance lines provides an
serious problem for their modelling. We can use for their wings modelling
the traditional
approach based on collisional interactions between atoms K and Na and
H, He and molecule only for qualitative analysis \cite{_UC2000}.

More sophisticated approaches based on quantum-chemical
consideration of the impact of potential fields provided by
different species on levels K and Na were proposed  recently by different
groups (see \cite{_BV2003} and \cite{_A2003}).

On the other hand, in  atmospheres of L-dwarfs the dust absorbs/scatters photons
 in
wide spectral spectral regions. Dust opacity affects the overall spectral
distributions (see \cite{_P2000} for more details). Perhaps, for
core and near wings of resonance lines K I and Na I 
we can still use the collisional approach \cite{_P2004}.

\medskip
\noindent {\small {\bf Infrared spectra: \hho bands}}
\medskip

Water bands cover the wide spectral regions in the infrared
spectra of UC dwarfs (see \cite{_P2002a} and the poster by
Lyubchik et al. on this session). For a long time the computation
of the most complete lists of \hho is the real challenge for
theoretical physics (see a review in \cite{_P2002}). In general,
incompleteness of water line lists used for the numerical analysis
of infrared spectra of UC dwarfs can increase our problems of
stellar spectra computations in different ways:

-- outer layers of model atmospheres computed with incomplete line
lists of \hho are ``too hot''.

-- results of spectral synthesis can be affected by incompleteness
of \hho lists.

Water bands in the IR are of interest for different topics.
Infrared CO band at 2.3 and 4.5 micron can be used for
determination of basic parameters of UC dwarfs: abundances,
effective temperatures, rotational velocities (see
\cite{_Jones2002}, \cite{_P2002a}). For their theoretical
modelling the use of reliable list of \hho lines is of crucial
importance (see \cite{_Jones2004} for more details).

\medskip
\noindent {\small {\bf Lithium test}}
\medskip

``Lithium test'' was proposed Rebolo et al. \cite{_Rebolo1992} to
identify brown dwarfs from the population of LMS. Before 1995 L-
and T-dwarfs were not known, and main attention was paid for the
low-gravity M-dwarfs. They suggested that at least part of low-mass 
dwarfs in young open
clusters should preserve their lithium. Observation of lithium lines in
spectra of late M-dwarfs  provides the direct evidence of their
substellar nature. 
Pavlenko et al. \cite{_P1995} showed that lithium lines can be
detected in spectra of brown dwarfs despite of  severe blending of the
atomic lines by molecular bands. Later lithium lines were really
found in spectra of some brown dwarfs (Teide1 \cite{_Rebolo1996},
Kelu1\cite{_Ruiz1998}, etc. see \cite{_Basri2000}).

\begin{figure}
\centerline{\includegraphics[width=8.0cm]{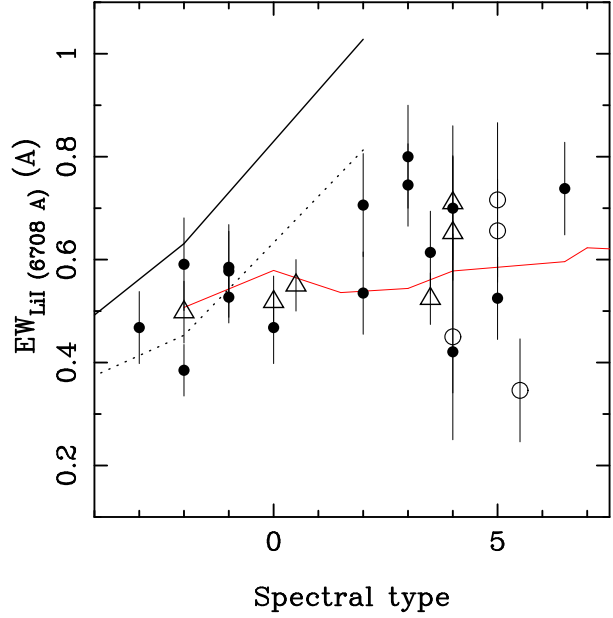}}
\caption{Comparison of pseudoequivalent widths (pEW) of the
lithium resonance doublet lines 670.8 nm computed for log N(Li) = 3.2
and observed in spectra in young
dwarfs of $\sigma$ Ori cluster.
TiO line list by Plez
\cite{_Plez1998} and NextGen model atmospheres\cite{_NextGen} of
solar metallicity \cite{_AG} were used in theoretical computations.
Solid and dashed lines in the left part of the plot indicate the
conventional curves of growth for the line
computed for log N(Li) = 3.2 and 2.0,
respectively.
Open circles and open triangles indicate
sources with H$_{\alpha}$ emission of pEW $>$ 1 nm and objects with forbidden
emission lines, respectively --
see Zapatero Osorio ~ \etal
\cite{_ZO2002} for more details. \label{_LiSOri}}
\end{figure}

On the other hand, observation of lithium lines in spectra of
late-type low gravity dwarfs of open clusters provide the
information about their age. Due to theoretical predictions 
(see refs. in \cite{_ZO2002}) the
smallest objects should be cooled very quickly, 
i.e within time scales of a few Myrs. Still
young, i.e low gravity dwarfs of ages 3-5 Myrs preserve their
lithium as well. In Fig. \ref{_LiSOri} results of determination of
litium abundances  in atmospheres of the low-mass dwarfs of open
cluster $\sigma$ Ori are showed. Note, these results are based on
analysis of pseudequivalent widths of lithium lines (see \cite
{_P1997} and \cite{_ZO2002} for more details) -- measurements of
the pseudoequivalent widthes are provided in respect to the local
pseudocontinuum formed by molecular lines.

Perhaps, determination of masses of brown dwarfs is the main
problem. Fortunately, often brown dwarfs form binary systems. The
study of the low-mass objects  is of special interest. First
observations of GJ569B provide some evidences about its substellar
nature (see refs in \cite{_ZO2001}. Still, later observations of
\Martin ~ \etal \cite{_Martin2000} on Keck Telescope show that GJ569B is double
system -- GJ569Ba and GJ569Bb are orbiting with period 892 $\pm$ 25  days
\cite{_Lane2001}.
Lithium test for this system is of crutial importance. However, in
this case we should manage a combine spectrum formed in
atmospheres of both components of different masses.

Later the application of the ``lithium test'' was discussed for
L-dwarfs and even T-dwarfs (see \cite{_P2000}). Indeed, lithium
lines were observed in spectra of some UC dwarfs.

%%%%%%%%%%%%%%%%%%%%%%%%%%%%%%%%%%%%%%%%%%
%%%%%%%%%%%%%%%%%%%%%%%%%%%%%%%%%%%%%%%%%%

\medskip
\noindent {\small {\bf Deuterim test}}
\medskip

In the cores of the ultracool dwarfs the correlations effects
between ions dominate lowering Coloumb barrier between particles
(see \cite{_Chabrier2000a} for more details).Still temperatures in
the interiors of UC dwarfs of masses M $<$ 13M$_J$ cannot be high
enough (T $<$ 0.5 MK) to initiate there a nuclear burning of
deuterium.

\Bejar ~ \etal \cite{_Bejar1999} propose to use observations of
lines of deuterium contained species to determine the ages/masses
of the smallest UC dwarfs. The task is very difficult in both
theoretical and observational aspects. The simplest case would be
proposed consists of the analysis of HDO/H$_2$O spectra in the IR
spectra of UC dwarfs . Still, HDO lines are very blended by \hho
lines \cite{_P2002}. From one side, we should have very accurate
line lists both \hho and HDO. Observed intensities of HDO lines
cannot exceed a few percent (see ibid and \cite{_Chabrier2000a}).
Moreover, IR spectrum of UC dwarfs should contain lines of other
polyatomic species (CH$_4$ and others). These factors increase the
demands for a capacity of observational facilities and the quality of
theoretical data to identify and to carry the analysis of HDO
lines in spectra of UC dwarfs.

%%%%%%%%%%%%%%%%%%%%%%%%%%%%%%%%%%%%%%%%%%
\begin{figure}
\centerline{\includegraphics[width=8.0cm]{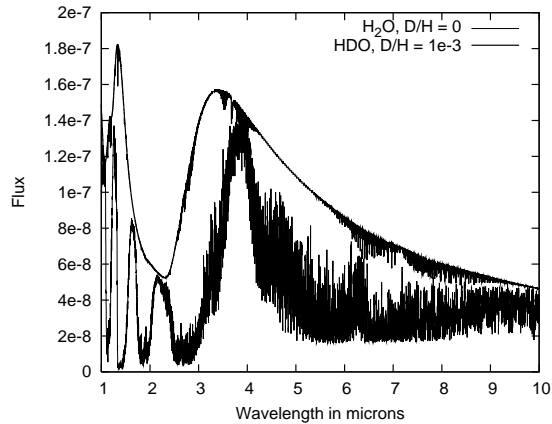}}
\caption{Computed spectra of \hho and HDO for different ratios D/H.
Computations were carried for model atmosphere 1200/5.0 by Tsuji (1998), AMES
line lists \cite{_PS1998}, with step 0.5 \AA, see \cite{_P2002}
for more details.
\label{_HDO}}
\end{figure}
%%%%%%%%%%%%%%%%%%%%%%%%%%%%%%%%%%%%%%%%%%

\bigskip
\noindent {\small {\bf ACKNOWLEDGEMENTS}}
\medskip
I thank Maria Rosa Zapatero Osorio (LAEFF, Spain) for  her highlighted and
helpful comments. I am grateful my collaborators and coauthors
Hugh R.A. Jones (Univ. of Hertfordshire, UK),
 Rafael Rebolo, \Martin Eduardo, V\'{\i}ctor J. S. B\'ejar
(IAC, Spain) for their contribution. 

My inverstigations were supported by Small Recearch Grant  of AAS
and Royal Society travel grant and travel grants from Liverpool
University (UK). \noindent

%%%%%%%%%%%%%%%%%%%%%%%%%%%%%%%%%%%%%%%%%

%%%%%%%%%%%%%%%%%%%%%%%%%%%%%%%%%%%%%%%%%%

%%%%%%%%%%%%%%%%%%%%%%%%%%%%%%%%%%%%%%%%%%
\end{document}